\documentclass[article,twocolumn,prd,superscriptaddress,preprintnumbers,nofootinbib]{revtex4-1}
\usepackage[utf8]{inputenc}
\usepackage[normalem]{ulem}
\usepackage{amsmath,amsthm,amssymb,mathtools,physics,amsfonts,cancel,mhchem,tikz}
\usepackage{graphicx,flexisym,float,mwe}
\usepackage{comment} 
\usepackage{float}
\usepackage{placeins}
\usepackage[T1]{fontenc}
\usepackage[hidelinks]{hyperref}
\hypersetup{colorlinks=true,
    linkcolor=blue,
    filecolor=blue,
    urlcolor=blue,
    citecolor=blue,
    }

\begin{document}
\hfill \preprint{MI-HET-814}

\title{Short Baseline Neutrino Anomalies at Stopped Pion Experiments}

\author{Iain A. Bisset}
\email{ibisset@tamu.edu}

\author{Bhaskar Dutta}
\email{dutta@physics.tamu.edu}

\author{Wei-Chih Huang}
\email{s104021230@tamu.edu}

\author{Louis E. Strigari}
\email{strigari@tamu.edu}
\affiliation{Mitchell Institute for Fundamental Physics and Astronomy$,$ Department~ of ~Physics ~ and~ Astronomy$,$\\ Texas A$\&$M University$,$~College~ Station$,$ ~Texas ~77843$,$~ USA}

\begin{abstract}
Stopped-pion experiments that measure coherent elastic neutrino-nucleus scattering (CE$\nu$NS) are sensitive to sterile neutrinos via disappearance. Using timing and energy spectra to perform flavor decomposition, we show that the delayed electron neutrino component provides an independent test of short-baseline anomalies that hint at $\sim$ eV-mass sterile neutrinos. Dedicated experiments will be sensitive to nearly the entire sterile neutrino parameter space consistent with short-baseline data. 
\end{abstract}

\maketitle

\section{Introduction}
\par Several neutrino short-baseline (SBL), reactor, and pion-decay experiments may be interpreted as evidence for sterile neutrinos~\citep{Acero:2022wqg}. Pion-decay experiments include the Liquid Scintillator Neutrino Detector (LSND)~\citep{LSND:2001aii} and the Booster Neutrino Experiments (MiniBooNE and MicroBooNE)~\citep{MiniBooNE:2020pnu,MicroBooNE:2022sdp}, which searched for sterile neutrinos via appearance and disappearance. SBL results include those from the recent Baksan Experiment on Sterile Transitions (BEST)~\citep{Barinov:2022wfh}. Combining the BEST results with previous results from SAGE~\citep{SAGE:2009eeu} and GALLEX~\citep{GNO:2005bds} implies evidence for an electron neutrino ($\nu_{e}$) disappearance, corresponding to a signal in the 3+1 sterile neutrino parameter space at 4$\sigma$ confidence~\citep{Giunti:2022btk,Elliott:2023cvh}. These results are consistent with a global analysis of reactor neutrino data~\citep{Berryman:2020agd}. 

\par COHERENT is a stopped-pion experiment that searches for the neutral current coherent elastic neutrino-nucleus scattering (CE$\nu$NS) process~\citep{Akimov:2022oyb}. COHERENT has measured CE$\nu$NS via a series of experiments using a 1 GeV proton beam~\citep{COHERENT:2020iec,COHERENT:2021xmm}. The resulting neutrinos arrive in prompt ($\nu_{\mu}$) and delayed ($\nu_{e}$, $\bar{\nu}_{\mu}$) windows, providing a means to differentiate the flavor components by incorporating the observed recoil times of CE$\nu$NS events along with their energies. Several other experimental efforts with similar proton beam energies and baselines are attempting to measure CE$\nu$NS at stopped-pion experiments, including Coherent CAPTAIN-Mills (CCM) at Los Alamos National Laboratory~\citep{CCM:2021leg}, JSNS$^2$ at JPARC~\citep{JSNS2:2021hyk}, and the European Spallation Source~\citep{Abele:2022iml}. 

\par The energy and the baselines of the stopped pion experiments imply that they are able to probe the parameter space consistent with the LSND/MiniBooNE anomalies~\citep{Anderson:2012pn,Blanco:2019vyp,Alonso-Gonzalez:2023tgm}. In this paper, we examine the prospects for extending the sensitivity of stopped-pion experiments into the parameter space consistent with aforementioned SBL anomalies that provide hints for sterile neutrinos with mass scale $\gtrsim 1$ eV. We take advantage of the enhanced statistical analysis enabled by the timing information to perform flavor decomposition~\citep{Dutta:2019eml,Giunti:2019xpr}, and from this decomposition generate constraints in the 3+1 sterile parameter space using the existing COHERENT CE$\nu$NS data and projected future data sets. 

\par We demonstrate that the COHERENT data has started to probe space consistent with the SBL anomalies. Further, nearly the entire parameter space consistent with the SBL anomalies can be tested by the ongoing and proposed experiments, for example, COHERENT experiment with NaI~\citep{Akimov:2022oyb}, or the proposed CCM  implementation of a ton-scale CsI~\cite{CCM_CSI}. This shows that stopped pion experiments provide an important new method to investigate the existing anomalies, which bypasses systematics associated with existing SBL data.

\section{CE$\nu$NS AND STERILE NEUTRINOS}

\par We utilize the CE$\nu$NS channel to search for sterile neutrinos via neutrino disappearance. In the Standard Model (SM), the CE$\nu$NS cross section for scattering with a nucleus is
\begin{equation}
    \label{cevns-cross}
    \frac{d\sigma}{dE} = \frac{G_{F}^{2}Q_{V}^{2}m}{2\pi} F^{2}(q^{2}) \left[ 1 - \left( \frac{m E}{E_{\nu}^{2}} \right ) + \left( 1 - \frac{E}{E_{\nu}} \right)^{2} \right]
\end{equation}
where $m$ is the total mass of the nucleus, $E$ is the kinetic nuclear recoil energy, $E_{\nu}$ is the incident neutrino energy, $Q_V=(\frac{1}{2}-2 \sin^2\theta_W)Z - \frac{1}{2}N$ with proton number $Z$, neutron number $N$, and $F(q^{2})$ is taken to be the Helm form factor, given by
\begin{equation}
    \label{Helm}
    F(q^{2}) = 3\frac{j_{1}(qR_0)}{qR_0}e^{-q^{2}s^{2}/2}
\end{equation}
where $R_0$ is the diffraction radius and $s$ is the nuclear skin thickness, with $s = 0.9$ fm. The parameter $q$ is the exchanged 3-momentum magnitude, which is taken in the relativistic limit to be $\sqrt{2m E}$. The diffraction radius $R_0$ is given by 
\begin{equation}
    \label{R}
    R_0 = \sqrt{(1.23 A^{1/3} - 0.6)^2\textrm{fm}^2 + \frac{7}{3}\pi^2r_{0}^{2} - 5s^2}
\end{equation}
with $r_{0} = 0.52$ fm and $A$ being the nucleon number~\citep{Lewin:1996}. The neutron rms radius $R_n$ can then be derived via $R_n^2 = \frac{3}{5}R_0^2 + 3s^2$. The derived neutron rms radius for Cs in this methodology is $\approx$ 4.8 fm.~For Ar, this value is $\approx$ 3.4 fm~\cite{AristizabalSierra:2019zmy}. The CsI result is consistent with the analysis of COHERENT data, which gives $5.5_{-1.1}^{+0.9}$ fm~\cite{Cadeddu:2017etk,DeRomeri:2022twg}.  

\indent To include the effect of sterile neutrinos, we employ a 3+1 sterile neutrino model. This model is parameterized by the squared mass difference, $\Delta m_{41}^{2} = m_{4}^{2} - m_{1}^{2}$, where $m_4$ is the mass of the sterile state and $m_1$ represents any of the active states. In addition there are the three squared mixing matrix elements $|U_{e4}|^{2}$, $|U_{\mu4}|^{2}$, and $|U_{\tau4}|^{2}$, with $|U_{s4}|^{2}$ fixed by unitarity. Since COHERENT is only sensitive to electron and muon neutrinos, we simply take $|U_{\tau4}|^{2} = 0$. We take the short-baseline approximation of the oscillation probabilities between flavor states $\alpha$ and $\beta$, which are given by
\begin{equation}
    \label{sterile-Prob}
    P_{\alpha \beta}(E_{\nu}) = \left| \delta_{\alpha \beta} - \sin^{2}{2 \theta_{\alpha \beta}}\sin^{2}{\frac{\Delta m_{41}^{2}L}{4E_{\nu}}} \right|
\end{equation}
with the expression $\sin^{2}{2 \theta_{\alpha \beta}}$ defined in terms of the matrix elements via
\begin{equation}
    \label{angles}
    \sin^{2}{2 \theta_{\alpha \beta}} = 4\left| \delta_{\alpha \beta}U_{\alpha 4}U_{\beta 4}^{*} - |U_{\alpha 4}|^{2} |U_{\beta 4}|^{2} \right|.
\end{equation}
The squared mass differences between active neutrinos are taken to be approximately zero and we have used $\Delta m_{42}^{2} \approx \Delta m_{41}^{2}$. 

\section{Prior Experiments}

\par BEST utilizes a $^{51}$Cr source of $\nu_{e}$ with energies in the range 437-752 keV, which then induce conversion of $^{71}$Ga to $^{71}$Ge via inverse beta decay. The experiment is sensitive to sterile neutrinos via $\nu_{e}$ disappearance, and as such is sensitive to the parameters $|U_{e4}|^2$ and $\Delta m_{41}^{2}$. Comparing the Ge production to that expected assuming pure SM, a deficit was observed. The $\nu_{e}$ at BEST had an average path length through the Ga target of $\approx$ 0.53m. Similarly, the SAGE and GALLEX experiments had average path lengths of $\approx$ 0.73m and $\approx$ 1.9m, respectively~\citep{Barinov:2022wfh,SAGE:1999,GALLEX:1998}. 

\par At longer baseline, the LSND experiment was an appearance experiment sensitive to $\bar{\nu}_{\mu} \rightarrow \bar{\nu}_{e}$ and $\nu_{\mu} \rightarrow \nu_{e}$ oscillations, allowing constraints to be established on the parameters $\sin^{2}{2 \theta_{\mu e}}$ and $\Delta m_{41}^{2}$.  Signals of oscillation to $\bar{\nu}_{e}$ are reconstructed from the coincidence of a positron from $\bar{\nu}_{e}p \rightarrow e^{+}n$ and a 2.2 MeV photon from the subsequent reaction $np \rightarrow d\gamma$. Here we compare to the results from the $\bar{\nu}_{\mu} \rightarrow \bar{\nu}_{e}$ analysis, which makes use of $\bar{\nu}_{\mu}$ resulting primarily from $\mu^+$ decay-at-rest. Relative to the SM expectation, LSND reported an excess with a 3.8$\sigma$ confidence~\citep{LSND:2001aii}.

\par Similar to LSND, MiniBooNE is an appearance experiment using 8 GeV protons on beryllium to produce decay-in-flight neutrinos. MiniBooNE is sensitive to $\bar{\nu}_{\mu} \rightarrow \bar{\nu}_{e}$ and $\nu_{\mu} \rightarrow \nu_{e}$ oscillations via observations of $\bar{\nu}_{e}p \rightarrow e^{+}n$ and $\nu_{e}n \rightarrow e^{-}p$. We use constraints established from a combined analysis of these two modes. The results indicate an excess with confidence $4.8\sigma$ in the parameter space of $\sin^{2}{2 \theta_{\mu e}}$ and $\Delta m_{41}^{2}$~\citep{MiniBooNE:2020pnu}. Data from MicroBooNE can also be used to investigate the $\nu_{\mu}$ and $\nu_{e}$ disappearance channels~\citep{MicroBooNE:2022sdp,Denton:2021czb,Arguelles:2022}. This, for example, leads to an upper bound on $|U_{\mu 4}|^2$ of $\approx$ 0.02 for a squared mass difference of 1 $\textrm{eV}^2$~\citep{Arguelles:2022}. For the purpose of this work, we simply compare to the appearance channel result from MiniBooNE.

\section{Data Analysis}

\indent COHERENT use a CsI target with mass of 14.6 kg and an exposure time of 587.65 days, with the detector at a distance of $L = 19.3$m from the neutrino source. The beam produced $N_{\textrm{POT}} = 3.198 \times 10^{23}$ protons-on-target (POT) impinging on Hg, with an average energy of 0.984 GeV, and an expected neutrino production rate per flavor of $r \approx 0.0848$ per POT~\citep{COHERENT:2021xmm}. COHERENT has presented results in the form of observed nuclear recoil events classified by their energies in photoelectrons (PE), and their corresponding recoil times. Data is given for both beam-on and beam-off periods, with the beam-off data used as the steady-state background (SSBG) in our analysis. Additional beam-related background events corresponding to beam-related neutron (BRN) and neutrino-induced neutron (NIN) events are also included~\citep{COHERENT:2021xmm}. 

\par COHERENT has also presented data for a CE$\nu$NS detection with liquid Ar (LAr). This experiment was performed with a fiducial detector mass of 24.4 kg, an exposure time of 229.95 days, and a baseline of 27.5m. The experiment had $N_{POT} = 1.38 \times 10^{23}$ protons-on-target with a neutrino production rate per flavor of $r \approx 0.09$ per POT~\citep{COHERENT:2020iec}. 

\par For our analysis of the CsI data, we bin the data in 9 PE bins and 11 timing bins in the same manner as presented by COHERENT in the ranges 0-60 PE and 0-6 $\mu$s~\citep{COHERENT:2021xmm}. Conversion from PE to kinetic recoil energy in keV is performed via the quenched recoil light yield of 13.35 PE/$\textrm{keV}_{\textrm{ee}}$ and the measured quenching factor. For the LAr case, we use the same binning as the data released by COHERENT. Energy resolution and efficiency are accounted for through the models provided by COHERENT~\citep{COHERENT:2020iec,COHERENT:2021xmm}. 

\begin{figure}[h]
    \centering
    \includegraphics[width=\columnwidth]{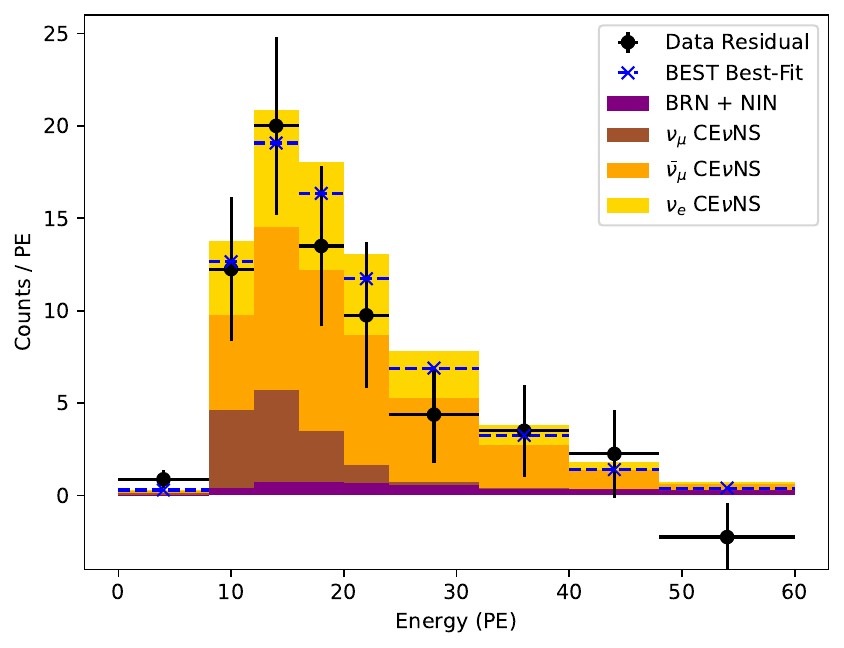}
    \caption{Predicted and measured energy spectrum, measured in photoelectrons (PE), for the CsI run at COHERENT. Data points with errors are the residuals after the SSBG has been subtracted. The stacked shaded histograms show the contributions from the different flavors and the beam related backgrounds. The dashed histogram provides the prediction for the best fit parameters as deduced from the BEST experiment.}
    \label{fig:spectra}
\end{figure}

\indent Given these definitions, the analysis proceeds by calculating the predicted number of total CE$\nu$NS events in each bin. The number of CE$\nu$NS events predicted in the jth energy bin and kth timing bin for CsI is given by
\begin{multline}
     \label{N-Energy}
    N_{jk}^{\textrm{CE$\nu$NS}} = \sum_{\alpha} \sum_{\substack{\beta = \textrm{Cs, I}\\\gamma = \nu_{\mu}, \bar{\nu}_{\mu}, \nu_{e}}} \mathcal{N}_{\beta} \int \int \int \int dtdE_{rec}dE_{true}dE_{\gamma} \\ \times \epsilon_{T}(t)\epsilon_{E}(E_{rec}) P_{E}(E_{rec}, E_{true}) P_{\gamma \alpha}(E_{\gamma}) \\ \times f_{T, \gamma}(t) f_{E, \, \gamma}(E_{\gamma}) \frac{d\sigma_{\beta}}{dE_{true}}(E_{true}, E_{\gamma})
\end{multline}
where the constant $\mathcal{N}_{\beta}$ is the total number of nuclei of type $\beta$ in the detector and $P_{\gamma \alpha}(E_{\gamma})$ is the oscillation probability for a neutrino of type $\gamma$ to have oscillated to type $\alpha$ at the detector. The sum over $\alpha$ occurs over each neutrino type. The integral over the reconstructed energy $E_{rec}$ goes over the jth bin, the integral over the true energy $E_{true}$ goes from 0 to the maximum allowed recoil energy of $\approx 2E_{\gamma}^2 / m$, and the integral over the incident neutrino energy $E_{\gamma}$ goes from the minimum neutrino energy of $\approx \sqrt{\frac{mE_{true}}{2}}$ to the maximum neutrino energy of $m_{\mu} / 2$. The value of $E_{\gamma}$ used in the upper bound of the $E_{true}$ integration is approximated by $m_{\mu} / 2$. The integral over time occurs over the kth timing bin. 

\indent The efficiencies in energy and time are accounted for by the functions $\epsilon_{E}$ and $\epsilon_{T}$, while energy resolution is accounted for by the function $P_{E}$. The functions $f_{E, \gamma}$ and $f_{T, \gamma}$ account for the distributions of the neutrino flux in energy and time, with $f_{T, \gamma}$ normalized to unity and $f_{E, \gamma}$ normalized to the total number of incident neutrinos of flavor $\gamma$. A SM calculation for the CsI case using our methodology is shown in energy space in Figure \ref{fig:spectra} along with the data residual for comparison. Also shown are contributions from the different flavor components, and for additional comparison the prediction for the best-fitting BEST sterile neutrino model. 

\indent To establish statistical constraints on the sterile neutrino parameter space, we perform a Bayesian analysis by first defining a likelihood $\mathcal{L}(\Vec{\theta})$. The vector $\Vec{\theta}$ represents the sterile neutrino parameters. For uncertainties present in the CsI run, $\alpha$ and $\beta$ are nuisance parameters serving to represent gaussian systematic uncertainties in CE$\nu$NS production and SSBG. To be in line with the reported uncertainty from COHERENT, we take $\sigma_{\alpha} = 0.1146$, accounting for the sum in quadrature of a 10\% flux uncertainty, 3.8\% quenching factor uncertainty, and 4.1\% energy efficiency uncertainty. The background uncertainty is taken to be $\sigma_{\beta} = 0.021$~\citep{COHERENT:2021xmm}. For the LAr analysis, the two nuisance parameters are taken to modify CE$\nu$NS production and the source of background uncertainty that generates the largest variation in the predicted number of counts (prompt beam-related neutrons) with uncertainties of 13\% and 32\% respectively~\citep{COHERENT:2020iec}. 

\indent The mean value, $\lambda$, predicted by this analysis in each bin is then a sum of the predicted CE$\nu$NS events and the backgrounds. 
For the CsI analysis this mean is
\begin{equation}
    \label{lambda}
    \lambda_{jk} = ( 1 + \alpha)N_{jk}^{\textrm{CE$\nu$NS}}(\Vec{\theta} ) + (1 + \beta)N_{jk}^{\textrm{SSBG}} + N_{jk}^{\textrm{BRN}} + N_{jk}^{\textrm{NIN}}
\end{equation}
Each bin is then assigned a poisson likelihood, giving a total likelihood of 
\begin{multline}
    \label{likelihood}
    \mathcal{L}(\Vec{\theta}) = \prod_{(j,k)} \int \int \textrm{exp}\{-\lambda_{jk}\}\frac{\{\lambda_{jk}\}^{N^{obs}_{jk}}}{N^{obs}_{jk}!} \\ \times \frac{\textrm{exp}(-\alpha^{2}/2\sigma_{\alpha}^{2})}{\sqrt{2\pi \sigma_{\alpha}^{2}}} \frac{\textrm{exp}(-\beta^{2}/2\sigma_{\beta}^{2})}{\sqrt{2\pi \sigma_{\beta}^{2}}}d\alpha d\beta
\end{multline}
where $N^{obs}$ are the observed counts. We assume constant priors for the model parameters $\Vec{\theta}$. To account for uncertainty in the identification of the start time $t = 0$, our analysis of constraints from the current CsI data allows the predicted timing distribution to shift with an offset in the range -0.25 to 0.25 $\mu$s with a flat prior. The LAr likelihood is analogous to Equation \ref{likelihood}, but with a gaussian likelihood assigned to each bin and with an extra dimension in binning corresponding to the pulse shape discrimination parameter $F_{90}$. To arrive at a posterior probability distribution, we use the MultiNest software package \cite{Buchner:2014,Feroz:2007kg,Feroz:2013hea,Feroz:2008xx}.

\section{Results}
\par
\begin{figure}[h]
    \centering 
    \includegraphics[width = 0.475\textwidth]{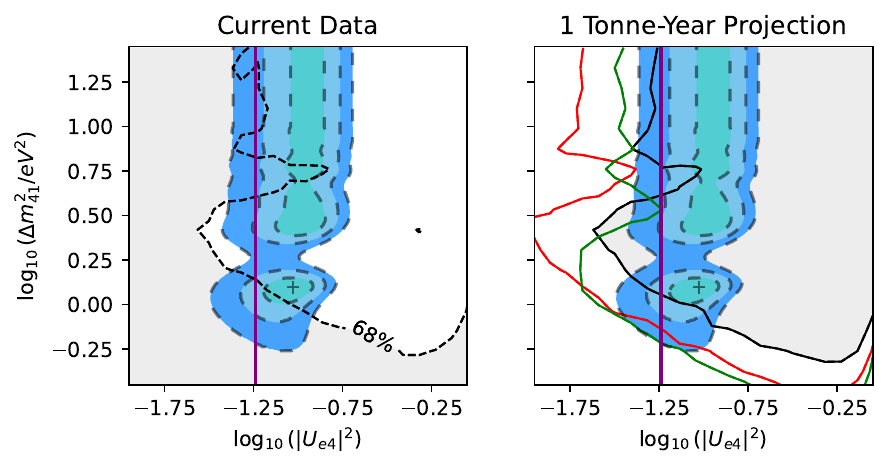}
    \caption{\label{fig:current} Allowed regions in the sterile parameter space. The left plot contains the 68\% level constraint from current CsI data, assuming $R_n = 4.8$ fm. The right plot shows 90\% exclusions from SM projections at 1 tonne-year with the same assumption for $R_n$, with black being CsI with current systematics, red being CsI with improved systematics, and green being LAr with improved systematics. The purple vertical line is the 95\% exclusion line from solar $\nu_{e}$~\citep{Goldhagen:2021kxe}. The blue and teal regions define the 1$\sigma$, 2$\sigma$, and 3$\sigma$ regions from BEST~\citep{Barinov:2022wfh}.}
\end{figure}

 We start by examining a two-dimensional parameter space of $\Delta m_{41}^2$ and $|U_{e4}|^2$, setting $|U_{\mu 4}|^2 = 0$. For a fixed value of $R_n = 4.8$ fm, the left panel of Figure~\ref{fig:current} shows the analysis of CsI data as compared to the region allowed by the combined SBL data. This shows that the majority of the SBL parameter space is still consistent with the COHERENT CsI data. 

\par To examine the future reach of CsI data, in the right panel of Figure~\ref{fig:current} we show the sensitivity for a SM projection at an exposure of 1 tonne-yr. Here we have simply scaled the backgrounds from the COHERENT data to the assumed exposure and take the same assumed value of $R_n = 4.8$ fm. To compare to the effects of improved systematics and background reduction, we show one case using the same uncertainties as reported by COHERENT and one case with uncertainty in CE$\nu$NS production reduced to 3\% and background uncertainty reduced to 1\%, with backgrounds reduced by a factor of 10 relative to the aforementioned scaling. Also shown are results for the LAr analysis, assuming the same improvement upon systematics and backgrounds. 

\par It is interesting to compare the left and right panels of Figure~\ref{fig:current}. In particular, we see that the region of the excluded parameter space shifts, so that the SBL data is allowed by current COHERENT data, but ruled out by future projections. This behavior is due to the assumed value of $R_n$ in the analysis of the COHERENT CsI data. Theoretical expectations place an upper bound on $R_n$ of $\approx$ 5.1 fm~\citep{AristizabalSierra:2019zmy}. For the purpose of this analysis, we take a conservative estimate with $R_n = 4.8$ fm. If we would, on the other hand, take $R_n = 5.5$ fm, we would see COHERENT exclusion curves that resemble more the right panel. This highlights the inherent degeneracy between $R_n = 5.5$ fm and the sterile parameters with the existing COHERENT data arising due to the uncertainty present at this level of exposure and systematics. This degeneracy can be broken with future data, which does not depend on the assumed $R_n$. We see that for a pure SM prediction for 1 tonne-year exposure, we have sensitivity to nearly the entire SBL parameter space using a different $\nu$ energy scale. Though our analysis in Figure~\ref{fig:current} assumes $U_{\mu 4} = 0$, we find similar results if we instead marginalize over this parameter. 

\par
\begin{figure}[h]
    \includegraphics[width = 0.475\textwidth]{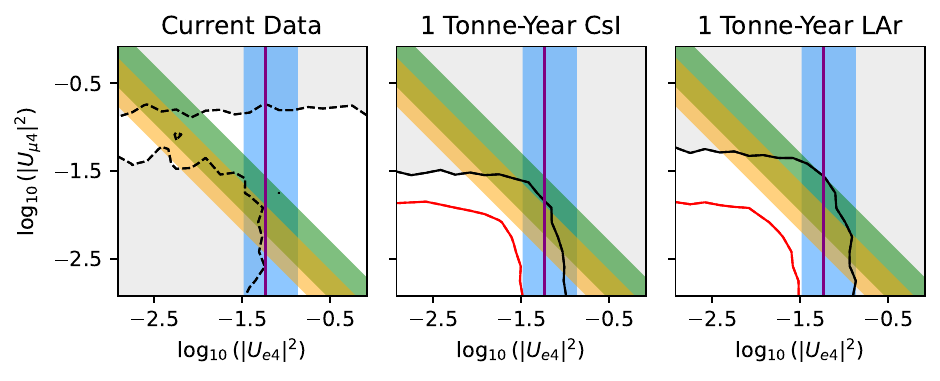} \\
    \includegraphics[width = 0.475\textwidth]{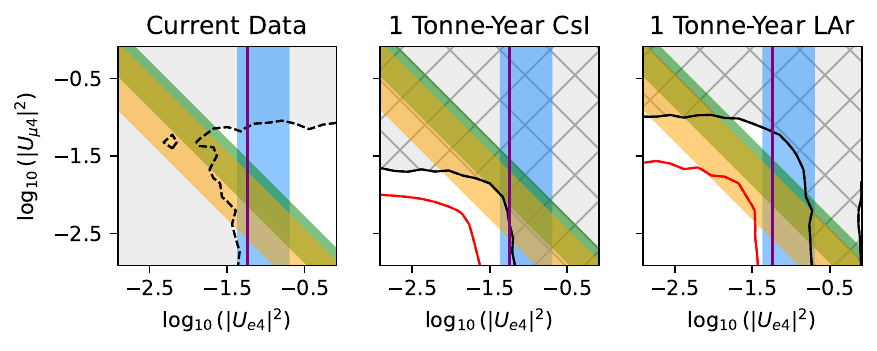}
    \caption{\label{fig:matrixspace} Allowed regions in the sterile parameter space for an assumed squared mass difference of $1$ eV$^2$ (top row) and $13$ eV$^2$ (bottom row). The left plot contains the 68\% level constraint from current CsI data. The right plots show 90\% exclusions from SM projections of CsI and LAr at 1 tonne-year, with black being with current systematics and red being with improved systematics. The purple vertical line is the 95\% exclusion line from solar $\nu_{e}$~\citep{Goldhagen:2021kxe}. The orange region defines the $4\sigma$ allowed space from MiniBooNE and the green region defines the 99\% allowed space from LSND~\citep{MiniBooNE:2020pnu,LSND:2001aii}. The blue band defines the $3\sigma$ allowed region from BEST~\citep{Barinov:2022wfh}.}
\end{figure}

 We now extend to consider a scenario in which $|U_{e4}|^2$, and $|U_{\mu4}|^2$ are allowed to be free. For assumed squared mass differences of $1 \, \textrm{eV}^2$ (top row) and $13 \, \textrm{eV}^2$ (bottom row), the first column of Figure~\ref{fig:matrixspace} shows the region consistent with the current CsI data. In this space, the SBL constraint is represented by a vertical band. This shows that nearly all of the region allowed by the SBL space is consistent with the current CsI data, similar to what is shown in Figure~\ref{fig:current}. The middle and right columns of Figure~\ref{fig:matrixspace} show the projected constraints for a 1 tonne-year exposure for CsI and LAr, respectively. We see that the constraints projected for CsI are stronger, because of the larger cross section and the reduced backgrounds in this case. 

\section{Discussion and conclusion}
\par Using information on the timing and energy spectra to separate the flavor components, we have used COHERENT data to constrain the parameter space consistent with the short baseline neutrino oscillation anomalies. We have focused in particular on the delayed $\nu_e$ component from stopped pion sources. We show that with $\sim 1$ tonne-year of exposure, and given current measured background levels and source flux uncertainties, stopped pion experiments will be able to cover nearly the entire mass and mixing parameter space associated with short baseline data from BEST, Gallium, LSND, and MiniBooNE. In addition to providing an independent check on SBL anomalies, the stopped-pion experiments will be able to test the SBL results at a different, higher characteristic energy scale. While MicroBoone constraints~\cite{MicroBooNE:2022sdp} are able to test part of the SBL sterile neutrino parameter space for non-zero $U_{e4}$ and $U_{\mu 4}$, stopped-pion experiments are able to cover nearly all of the space for these assumptions. 

\par Forthcoming experiments will likely be able to reduce the systematics to a level below which we have considered. For example, with a segmented $\sim$ ton-scale CsI detector, detector background levels will be reduced relative to those measured by COHERENT. Even with current background levels, forthcoming data from NaI at COHERENT will be able to establish important new constraints. Reducing the uncertainty on the source flux below the $\sim 11\%$ will also improve the results we discuss. Source flux uncertainty may be reduced via e.g. a D2O calibration as is being pursued by COHERENT~\citep{Akimov:2022oyb}, or even by placing detectors at different locations along the beamline~\citep{Anderson:2012pn}. 

\begin{acknowledgments}
We thank K. Kelly, W. Louis, R. Mahapatra, D. Pershey, and R. Van de Water for discussions on this paper. B.~D. and L.~E.~S. are supported by the DOE Grant No. DE-SC0010813. We acknowledge support from the Texas A\&M University System National Laboratories Office and Los Alamos National Laboratory.
\end{acknowledgments}

\bibliographystyle{bibi}
\bibliography{refs}

\providecommand{\href}[2]{#2}\begingroup\raggedright\begin{thebibliography}{10}

\bibitem{Acero:2022wqg}
M.~A. Acero et~al., \emph{{White Paper on Light Sterile Neutrino Searches and Related Phenomenology}},  \href{https://arxiv.org/abs/2203.07323}{{\ttfamily 2203.07323}}.

\bibitem{LSND:2001aii}
{\scshape LSND} Collaboration, A.~Aguilar et~al., \emph{{Evidence for neutrino oscillations from the observation of $\bar{\nu}_e$ appearance in a $\bar{\nu}_\mu$ beam}}, \href{https://doi.org/10.1103/PhysRevD.64.112007}{\emph{Phys. Rev. D} {\bfseries 64} (2001) 112007} [\href{https://arxiv.org/abs/hep-ex/0104049}{{\ttfamily hep-ex/0104049}}].

\bibitem{MiniBooNE:2020pnu}
{\scshape MiniBooNE} Collaboration, A.~A. Aguilar-Arevalo et~al., \emph{{Updated MiniBooNE neutrino oscillation results with increased data and new background studies}}, \href{https://doi.org/10.1103/PhysRevD.103.052002}{\emph{Phys. Rev. D} {\bfseries 103} (2021) 052002} [\href{https://arxiv.org/abs/2006.16883}{{\ttfamily 2006.16883}}].

\bibitem{MicroBooNE:2022sdp}
{\scshape MicroBooNE} Collaboration, P.~Abratenko et~al., \emph{{First Constraints on Light Sterile Neutrino Oscillations from Combined Appearance and Disappearance Searches with the MicroBooNE Detector}}, \href{https://doi.org/10.1103/PhysRevLett.130.011801}{\emph{Phys. Rev. Lett.} {\bfseries 130} (2023) 011801} [\href{https://arxiv.org/abs/2210.10216}{{\ttfamily 2210.10216}}].

\bibitem{Barinov:2022wfh}
V.~V. Barinov et~al., \emph{{Search for electron-neutrino transitions to sterile states in the BEST experiment}}, \href{https://doi.org/10.1103/PhysRevC.105.065502}{\emph{Phys. Rev. C} {\bfseries 105} (2022) 065502} [\href{https://arxiv.org/abs/2201.07364}{{\ttfamily 2201.07364}}].

\bibitem{SAGE:2009eeu}
{\scshape SAGE} Collaboration, J.~N. Abdurashitov et~al., \emph{{Measurement of the solar neutrino capture rate with gallium metal. III: Results for the 2002--2007 data-taking period}}, \href{https://doi.org/10.1103/PhysRevC.80.015807}{\emph{Phys. Rev. C} {\bfseries 80} (2009) 015807} [\href{https://arxiv.org/abs/0901.2200}{{\ttfamily 0901.2200}}].

\bibitem{GNO:2005bds}
{\scshape GNO} Collaboration, M.~Altmann et~al., \emph{{Complete results for five years of GNO solar neutrino observations}}, \href{https://doi.org/10.1016/j.physletb.2005.04.068}{\emph{Phys. Lett. B} {\bfseries 616} (2005) 174} [\href{https://arxiv.org/abs/hep-ex/0504037}{{\ttfamily hep-ex/0504037}}].

\bibitem{Giunti:2022btk}
C.~Giunti, Y.~F. Li, C.~A. Ternes, O.~Tyagi and Z.~Xin, \emph{{Gallium Anomaly: critical view from the global picture of \ensuremath{\nu}$_{e}$ and $ {\overline{\nu}}_e $ disappearance}}, \href{https://doi.org/10.1007/JHEP10(2022)164}{\emph{JHEP} {\bfseries 10} (2022) 164} [\href{https://arxiv.org/abs/2209.00916}{{\ttfamily 2209.00916}}].

\bibitem{Elliott:2023cvh}
S.~R. Elliott, V.~Gavrin and W.~Haxton, \emph{{The Gallium Anomaly}},  \href{https://arxiv.org/abs/2306.03299}{{\ttfamily 2306.03299}}.

\bibitem{Berryman:2020agd}
J.~M. Berryman and P.~Huber, \emph{{Sterile Neutrinos and the Global Reactor Antineutrino Dataset}}, \href{https://doi.org/10.1007/JHEP01(2021)167}{\emph{JHEP} {\bfseries 01} (2021) 167} [\href{https://arxiv.org/abs/2005.01756}{{\ttfamily 2005.01756}}].

\bibitem{Akimov:2022oyb}
D.~Akimov et~al., \emph{{The COHERENT Experimental Program}},  in \emph{{Snowmass 2021}}, 4, 2022, \href{https://arxiv.org/abs/2204.04575}{{\ttfamily 2204.04575}}.

\bibitem{COHERENT:2020iec}
{\scshape COHERENT} Collaboration, D.~Akimov et~al., \emph{{First Measurement of Coherent Elastic Neutrino-Nucleus Scattering on Argon}}, \href{https://doi.org/10.1103/PhysRevLett.126.012002}{\emph{Phys. Rev. Lett.} {\bfseries 126} (2021) 012002} [\href{https://arxiv.org/abs/2003.10630}{{\ttfamily 2003.10630}}].

\bibitem{COHERENT:2021xmm}
{\scshape COHERENT} Collaboration, D.~Akimov et~al., \emph{{Measurement of the Coherent Elastic Neutrino-Nucleus Scattering Cross Section on CsI by COHERENT}}, \href{https://doi.org/10.1103/PhysRevLett.129.081801}{\emph{Phys. Rev. Lett.} {\bfseries 129} (2022) 081801} [\href{https://arxiv.org/abs/2110.07730}{{\ttfamily 2110.07730}}].

\bibitem{CCM:2021leg}
{\scshape CCM} Collaboration, A.~A. Aguilar-Arevalo et~al., \emph{{First dark matter search results from Coherent CAPTAIN-Mills}}, \href{https://doi.org/10.1103/PhysRevD.106.012001}{\emph{Phys. Rev. D} {\bfseries 106} (2022) 012001} [\href{https://arxiv.org/abs/2105.14020}{{\ttfamily 2105.14020}}].

\bibitem{JSNS2:2021hyk}
{\scshape JSNS2} Collaboration, S.~Ajimura et~al., \emph{{The JSNS2 detector}}, \href{https://doi.org/10.1016/j.nima.2021.165742}{\emph{Nucl. Instrum. Meth. A} {\bfseries 1014} (2021) 165742} [\href{https://arxiv.org/abs/2104.13169}{{\ttfamily 2104.13169}}].

\bibitem{Abele:2022iml}
H.~Abele et~al., \emph{{Particle Physics at the European Spallation Source}}, \href{https://doi.org/10.1016/j.physrep.2023.06.001}{\emph{Phys. Rept.} {\bfseries 1023} (2023) 1} [\href{https://arxiv.org/abs/2211.10396}{{\ttfamily 2211.10396}}].

\bibitem{Anderson:2012pn}
A.~J. Anderson, J.~M. Conrad, E.~Figueroa-Feliciano, C.~Ignarra, G.~Karagiorgi, K.~Scholberg, M.~H. Shaevitz and J.~Spitz, \emph{{Measuring Active-to-Sterile Neutrino Oscillations with Neutral Current Coherent Neutrino-Nucleus Scattering}}, \href{https://doi.org/10.1103/PhysRevD.86.013004}{\emph{Phys. Rev. D} {\bfseries 86} (2012) 013004} [\href{https://arxiv.org/abs/1201.3805}{{\ttfamily 1201.3805}}].

\bibitem{Blanco:2019vyp}
C.~Blanco, D.~Hooper and P.~Machado, \emph{{Constraining Sterile Neutrino Interpretations of the LSND and MiniBooNE Anomalies with Coherent Neutrino Scattering Experiments}}, \href{https://doi.org/10.1103/PhysRevD.101.075051}{\emph{Phys. Rev. D} {\bfseries 101} (2020) 075051} [\href{https://arxiv.org/abs/1901.08094}{{\ttfamily 1901.08094}}].

\bibitem{Alonso-Gonzalez:2023tgm}
D.~Alonso-Gonz\'alez, D.~W.~P. Amaral, A.~Bariego-Quintana, D.~Cerdeno and M.~d.~l. Rios, \emph{{Measuring the Sterile Neutrino Mass in Spallation Source and Direct Detection Experiments}},  \href{https://arxiv.org/abs/2307.05176}{{\ttfamily 2307.05176}}.

\bibitem{Dutta:2019eml}
B.~Dutta, S.~Liao, S.~Sinha and L.~E. Strigari, \emph{{Searching for Beyond the Standard Model Physics with COHERENT Energy and Timing Data}}, \href{https://doi.org/10.1103/PhysRevLett.123.061801}{\emph{Phys. Rev. Lett.} {\bfseries 123} (2019) 061801} [\href{https://arxiv.org/abs/1903.10666}{{\ttfamily 1903.10666}}].

\bibitem{Giunti:2019xpr}
C.~Giunti, \emph{{General COHERENT constraints on neutrino nonstandard interactions}}, \href{https://doi.org/10.1103/PhysRevD.101.035039}{\emph{Phys. Rev. D} {\bfseries 101} (2020) 035039} [\href{https://arxiv.org/abs/1909.00466}{{\ttfamily 1909.00466}}].

\bibitem{CCM_CSI}
R.~Van~de Water, \emph{Probing the dark sector with accelerators: New opportunities!},  https://www.int.washington.edu/program/schedule/1205, April 17-21, 2023.
\newblock Talk at Interplay of Nuclear, Neutrino and BSM Physics at Low-Energies, INT, Seattle.

\bibitem{Lewin:1996}
J.~Lewin and P.~Smith, \emph{Review of mathematics, numerical factors, and corrections for dark matter experiments based on elastic nuclear recoil}, \href{https://doi.org/10.1016/S0927-6505(96)00047-3}{\emph{Astroparticle Physics} {\bfseries 6} (1996) 87}.

\bibitem{AristizabalSierra:2019zmy}
D.~Aristizabal~Sierra, J.~Liao and D.~Marfatia, \emph{{Impact of form factor uncertainties on interpretations of coherent elastic neutrino-nucleus scattering data}}, \href{https://doi.org/10.1007/JHEP06(2019)141}{\emph{JHEP} {\bfseries 06} (2019) 141} [\href{https://arxiv.org/abs/1902.07398}{{\ttfamily 1902.07398}}].

\bibitem{Cadeddu:2017etk}
M.~Cadeddu, C.~Giunti, Y.~F. Li and Y.~Y. Zhang, \emph{{Average CsI neutron density distribution from COHERENT data}}, \href{https://doi.org/10.1103/PhysRevLett.120.072501}{\emph{Phys. Rev. Lett.} {\bfseries 120} (2018) 072501} [\href{https://arxiv.org/abs/1710.02730}{{\ttfamily 1710.02730}}].

\bibitem{DeRomeri:2022twg}
V.~De~Romeri, O.~G. Miranda, D.~K. Papoulias, G.~Sanchez~Garcia, M.~T\'ortola and J.~W.~F. Valle, \emph{{Physics implications of a combined analysis of COHERENT CsI and LAr data}}, \href{https://doi.org/10.1007/JHEP04(2023)035}{\emph{JHEP} {\bfseries 04} (2023) 035} [\href{https://arxiv.org/abs/2211.11905}{{\ttfamily 2211.11905}}].

\bibitem{SAGE:1999}
{\scshape SAGE} Collaboration, J.~N. Abdurashitov et~al., \emph{{Measurement of the response of a gallium metal solar neutrino experiment to neutrinos from a ${}^{51}\mathrm{Cr}$ source}}, \href{https://doi.org/10.1103/PhysRevC.59.2246}{\emph{Phys. Rev. C} {\bfseries 59} (1999) 2246} [\href{https://arxiv.org/abs/hep-ph/9803418}{{\ttfamily hep-ph/9803418}}].

\bibitem{GALLEX:1998}
{\scshape GALLEX} Collaboration, W.~Hampel et~al., \emph{{Final results of the 51Cr neutrino source experiments in GALLEX}}, \href{https://doi.org/10.1016/S0370-2693(97)01562-1}{\emph{Phys. Lett. B} {\bfseries 420} (1998) 114}.

\bibitem{Denton:2021czb}
P.~B. Denton, \emph{{Sterile Neutrino Search with MicroBooNE\textquoteright{}s Electron Neutrino Disappearance Data}}, \href{https://doi.org/10.1103/PhysRevLett.129.061801}{\emph{Phys. Rev. Lett.} {\bfseries 129} (2022) 061801} [\href{https://arxiv.org/abs/2111.05793}{{\ttfamily 2111.05793}}].

\bibitem{Arguelles:2022}
C.~A. Arg\"uelles, I.~Esteban, M.~Hostert, K.~J. Kelly, J.~Kopp, P.~A.~N. Machado, I.~Martinez-Soler and Y.~F. Perez-Gonzalez, \emph{{MicroBooNE and the ${\ensuremath{\nu}}_{e}$ Interpretation of the MiniBooNE Low-Energy Excess}}, \href{https://doi.org/10.1103/PhysRevLett.128.241802}{\emph{Phys. Rev. Lett.} {\bfseries 128} (2022) 241802} [\href{https://arxiv.org/abs/2111.10359}{{\ttfamily 2111.10359}}].

\bibitem{Buchner:2014}
J.~Buchner et~al., \emph{{X-ray spectral modelling of the AGN obscuring region in the CDFS: Bayesian model selection and catalogue}}, \href{https://doi.org/10.1051/0004-6361/201322971}{\emph{A\&A} {\bfseries 564} (2014) A125} [\href{https://arxiv.org/abs/1402.0004}{{\ttfamily 1402.0004}}].

\bibitem{Feroz:2007kg}
F.~Feroz and M.~P. Hobson, \emph{{Multimodal nested sampling: an efficient and robust alternative to MCMC methods for astronomical data analysis}}, \href{https://doi.org/10.1111/j.1365-2966.2007.12353.x}{\emph{Mon. Not. Roy. Astron. Soc.} {\bfseries 384} (2008) 449} [\href{https://arxiv.org/abs/0704.3704}{{\ttfamily 0704.3704}}].

\bibitem{Feroz:2013hea}
F.~Feroz, M.~P. Hobson, E.~Cameron and A.~N. Pettitt, \emph{{Importance Nested Sampling and the MultiNest Algorithm}}, \href{https://doi.org/10.21105/astro.1306.2144}{\emph{Open J. Astrophys.} {\bfseries 2} (2019) 10} [\href{https://arxiv.org/abs/1306.2144}{{\ttfamily 1306.2144}}].

\bibitem{Feroz:2008xx}
F.~Feroz, M.~P. Hobson and M.~Bridges, \emph{{MultiNest: an efficient and robust Bayesian inference tool for cosmology and particle physics}}, \href{https://doi.org/10.1111/j.1365-2966.2009.14548.x}{\emph{Mon. Not. Roy. Astron. Soc.} {\bfseries 398} (2009) 1601} [\href{https://arxiv.org/abs/0809.3437}{{\ttfamily 0809.3437}}].

\bibitem{Goldhagen:2021kxe}
K.~Goldhagen, M.~Maltoni, S.~E. Reichard and T.~Schwetz, \emph{{Testing sterile neutrino mixing with present and future solar neutrino data}}, \href{https://doi.org/10.1140/epjc/s10052-022-10052-2}{\emph{Eur. Phys. J. C} {\bfseries 82} (2022) 116} [\href{https://arxiv.org/abs/2109.14898}{{\ttfamily 2109.14898}}].

\end{thebibliography}\endgroup

\end{document}